\RequirePackage{fix-cm}
\documentclass[twocolumn,epjc3]{svjour3} \usepackage[T1]{fontenc}
\usepackage[utf8]{inputenc}
\RequirePackage{graphicx}
\RequirePackage[numbers,compress]{natbib}
\usepackage{amsmath}
\usepackage{amssymb}

\usepackage{hyperref}
\hypersetup{
    colorlinks=true,
    linkcolor=blue,
    filecolor=magenta,
    urlcolor=cyan,
    citecolor=cyan
}

\journalname{Eur. Phys. J. C}

\def\be{\begin{equation}}
\def\ee{\end{equation}}

\def\bea{\begin{eqnarray}}
\def\eea{\end{eqnarray}}

\begin{document}

\title{$\Lambda$CDM suitably embedded in $f(R)$ with a non-minimal coupling to matter}

\author{Mar\'ia Ortiz-Ba\~nos\thanksref{e1,addr1}
        \and
        Mariam Bouhmadi-L\'opez\thanksref{e2,addr1,addr2}
        \and
        Ruth Lazkoz\thanksref{e3,addr1}
        \and
        Vincenzo Salzano\thanksref{e4,addr3} 
}

\thankstext{e1}{e-mail: maria.ortiz@ehu.eus}
\thankstext{e2}{e-mail: mariam.bouhmadi@ehu.eus}
\thankstext{e3}{e-mail: ruth.lazkoz@ehu.eus}
\thankstext{e4}{e-mail: vincenzo.salzano@usz.edu.pl}

\institute{Department of Physics, University of the Basque Country UPV/EHU, P.O. Box 644, 48080 Bilbao, Spain \label{addr1}
\and
IKERBASQUE, Basque Foundation for Science, 48011, Bilbao, Spain \label{addr2}
\and
Institute of Physics, University of Szczecin, Wielkopolska 15, 70-451 Szczecin, Poland \label{addr3}
}

\date{Received: date / Accepted: date}

\maketitle

\begin{abstract}
 In this work, we further study a metric modified theory of gravity which contains a non-minimal coupling to matter, more precisely, we assume two functions of the scalar curvature, $f_1$ and $f_2$, where the first one generalises the Hilbert-Einstein action, while the second couples to the matter Lagrangian. On the one hand, assuming a $\Lambda$CDM background, we calculate analytical solutions for the functions $f_1$ and $f_2$. We consider two setups: on the first one, we fix $f_2$ and compute $f_1$ and on the second one, we fix $f_1$ and compute $f_2$. Moreover, we do the analysis for two different energy density contents, a matter dominated universe and a general perfect fluid with a constant equation of state fuelling the universe expansion. On the other hand, we complete our study by performing a cosmographic analysis for $f_1$ and $f_2$. We conclude that the gravitational coupling to matter can drive the accelerated expansion of the universe.
\end{abstract}

\section{Introduction}

During the last decades modified theories of gravity have gathered a lot of attention due to their potential to explain certain cosmological phenomena  \cite{Nojiri:2006ri, Capozziello:2007ec, DeFelice:2010aj, Sotiriou:2008rp, Nojiri:2010wj, Capozziello:2011et, Clifton:2011jh}. These models arise as an alternative to General Relativity (GR) to explain the different events which occur throughout the whole evolution of the universe, from the primordial inflationary era to the recent speed up of the universe. For example, Starobinsky inflationary model \cite{1980PhLB} seats at the sweet spot of \textit{Planck} observations \cite{Akrami:2018odb}.
In addition, in what refers to the theory behind the late-time acceleration of the universe \cite{Riess:1998cb,Perlmutter:1998np,Perlmutter:2003kf}, there is no consensus on a definitive candidate, however, modified theories of gravity offer a big plethora of possibilities to describe the recent speed up of the universe.

The simplest approach to describe a modified gravity theory is within the framework of $f(R)$ metric gravity, which consists on generalising Hilbert-Einstein action to a general function of the scalar curvature, $R$ \cite{Capozziello:2007ec, DeFelice:2010aj, Sotiriou:2008rp}. If, in addition, we consider a coupling, though small, between the gravitational sector and matter, we would be considering what is known as $f(R)$ non-minimally coupled theories, first proposed in \cite{extraforce}.
 Those theories have gathered quite some attention since some years ago to describe different cosmological events.
 For a non-exhaustive but representative variety of publications on the topic please see \cite{Ramos:2017cot, Gomes:2016cwj, Bertolami:2008ab, Bertolami:2009ic, Bertolami_2010, Silva:2017weg, modifiedfried, Bertolami:2010cw, Ribeiro:2014sla, Bertolami:2013kca, Nesseris:2008mq, Bertolami:2017svl}. More precisely, in \cite{Ramos:2017cot}, gravitational baryogenesis is analysed finding compatible results with observations whereas in \cite{Gomes:2016cwj} some inflationary scenarios are studied and examples of observational predictions are given for some of the most common potentials in order to set limits on the scale of the non-minimal coupling. In \cite{Bertolami:2008ab}, the degeneracy of Lagrangian densities for a perfect fluid is discussed.  On the one hand, one of the motivations of analysing this kind of models is their ability to answer for the dark matter effects which are observed in our universe.
  In \cite{Bertolami:2009ic}, it is shown that with non-minimally coupled $f(R)$ theories it is possible to mimic known dark matter density profiles through a specific power-law coupling.
  In \cite{Bertolami_2010} it is shown how these theories can give rise to an additional contribution to the field equations, as compared with GR, that can play the role of dark matter. More recently in \cite{Silva:2017weg}, analytical solutions to the modified field equations are derived and, besides, constraints on the parameters of the model are obtained by comparing their predicted profiles for visible and dark matter with known ones.
 On the other hand,  it is still an unanswered question which gravitational theory could be behind the late-time acceleration of the universe, within this scope, $f(R)$ non-minimally coupled theories have been also studied as possible candidates for this accelerated behaviour as observed nowadays \cite{modifiedfried,  Bertolami:2010cw, Ribeiro:2014sla}. In \cite{Bertolami:2013kca} the authors study the evolution of the cosmological perturbations within this kind of theories and an analysis of the large-scale structures formed is carried. Matter density perturbations are also studied in \cite{Nesseris:2008mq}, where some constraints on a specific model are setted using the age of the oldest star clusters and primordial
nucleosynthesis bounds. Finally, in \cite{Bertolami:2017svl} it has been recently performed a gravitational waves analysis within these models.

In our work, we farther analyse $f(R)$ non-minimally coupled theories as a mean to succesfully describe the late-time acceleration of the universe.
We firstly construct two setups within the framework of \cite{extraforce} which mimic a $\Lambda$CDM universe. The action describing this theory is described by two functions, $f_1(R)$ and $f_2(R)$, the first one will play the role of a typical $f(R)$ metric theory and the second one is non-minimally coupled to matter through a small coupling constant, $\lambda$.
In the first setup, we fix $f_2(R)=R$ and compute $f_1(R)$ analytically, in the second one we reverse the roles, i.e. we fix $f_1(R)=R-2\kappa^2\Lambda$ and find an analytical solution for $f_2(R)$. Besides, we analyse each setup for two different energy-density contents and obtain the corresponding physical solutions analytically. Secondly, in order to get an order of magnitude of the parameters of the model for the two setups considered, we perform a cosmographic approach and we map it to the current observational values fitting $\Lambda$CDM \cite{Akrami:2018odb}. Moreover, we analyse in detail the physical meaning of the coupling parameter, $\lambda$, which, as we will see later, can be interpreted as a parameter intimately related to the late-time speed up of the universe.

The paper is organised as follows: In Sec.~\ref{sec:intro} we review $f(R)$ non-minimally coupled theories; in Sec.~\ref{sec:solutions} we calculate the analytical solutions for our theory in the two selected setups; in Sec.~\ref{sec:cosmogra} we perform a cosmographic analysis including a study of the effects of the coupling constant on the value of the cosmographic quantities and in Sec.~\ref{sec:results}, we sum up the obtained results.

\section{Non-minimally coupled \texorpdfstring{$f(R)$}{fR}}\label{sec:intro}

We consider a model which includes a non-minimal coupling between geometry and matter. The action reads \cite{extraforce}
\be\label{action}
 S=\int \left[\frac{f_1(R)}{2\kappa^2}+(1+\lambda f_2(R))\mathcal{L}\right]\sqrt{-g}~d^4x,
 \ee
where $f_i(R)~(i=1,2)$ are arbitrary functions of the Ricci scalar $R$, $g$ is the determinant of the metric $g_{\mu\nu}$,
$\kappa^2=8\pi G$, $\lambda$ is a coupling constant with length square units and $\mathcal{L}$ is the matter Lagrangian density.

\subsection{Field equations}

From the variation of the action (\ref{action}) with respect to the metric one gets the modified Einstein's equations \cite{extraforce}
\begin{eqnarray}\label{einstein}
F_1R_{\mu\nu}&-&\frac{1}{2}f_1g_{\mu\nu}-\left(\nabla_{\mu}\nabla_{\nu}-g_{\mu\nu}\square\right)F_1= \\
&-&2\lambda \kappa^2 F_2\mathcal{L}R_{\mu\nu}+2\lambda\kappa^2\left(\nabla_{\mu}\nabla_{\nu}-g_{\mu\nu}\square\right)\mathcal{L}F_2 \nonumber \\
&+&(1+\lambda f_2)\kappa^2T_{\mu\nu}, \nonumber
\end{eqnarray}
where $F_i(R)\equiv\frac{df_i(R)}{dR}~(i=1,2)$, $\square=g^{\mu\nu}\nabla_{\mu}\nabla_{\nu}$, $T_{\mu\nu}$ is the energy momentum tensor
and $R_{\mu\nu}$ is the Ricci tensor.
Taking the covariant derivative of Eq. (\ref{einstein}) one can deduce the following modified conservation equation for the energy momentum tensor \cite{extraforce}
\be\label{conservation}
\nabla^{\mu}T_{\mu\nu}^{}=\frac{\lambda F_2}{1+\lambda f_2}[g_{\mu\nu}\mathcal{L}-T_{\mu\nu}^{}]\nabla^{\mu}R.
\ee

\subsection{Dynamics of a homogeneous and isotropic Universe}

Considering a perfect fluid in a spatially flat FLRW scenario
\be
ds^2=-dt^2+a(t)^2\left[dr^2+r^2(d\theta^2+\sin^2\theta d\phi^2)\right],
\ee
one can check if the energy momentum tensor  is conserved or not by taking the $\mu=0$ component of Eq. (\ref{conservation}) (c.f. \cite{modifiedfried})
\be\label{cons}
\dot{\rho}+3H(P+\rho)=\frac{\lambda F_2}{1+\lambda f_2}[-\mathcal{L}-\rho]\dot{R},
\ee
where $\cdot\equiv d/dt$, $\rho$ is the matter density, $P$ is the pressure and $H=\dot{a}/a$.
From here, we note that if $\mathcal{L}=-\rho$, $\dot{\rho}$ is conserved. We will assume this is the case from now on.

As we are interested in computing the dynamics of our model, we take the $tt$ component of Eq. (\ref{einstein}) in order
to obtain the Friedmann equation \cite{modifiedfried}
\be\label{fried}
3H^2=\tilde{\kappa}^2(\rho+\rho_{f_1}+\rho_{f_2}),
\ee
where
\begin{align}
 \tilde{\kappa}^2&\equiv\frac{\kappa^2(1+\lambda f_2)}{F_1+2\lambda\kappa^2\mathcal{L}F_2},\\
 \rho_{f_1}&\equiv\frac{-6H\partial_tF_1+F_1 R-f_1}{2\kappa^2(1+\lambda f_2)},\\
 \rho_{f_2}&\equiv\frac{-6H\lambda\partial_t(\mathcal{L}F_2)+\lambda\mathcal{L}F_2R}{1+\lambda f_2}.
\end{align}
The theory reduces to GR for $f_1=R-2\Lambda\kappa^2$ and a vanishing $f_2$, while for a vanishing $f_2$
 we recover the standard metric $f(R)$ theory.
Here we have a second order differential equation on $f_1$ and $f_2$. We want to solve this equation
analytically, if possible. Our procedure will consist on
fixing a simple $f_1$ ( $f_2$ ) and solving for $f_2$ ( $f_1$ ).

\section{A \texorpdfstring{$\Lambda$}{Lambda}CDM geometry within non-minimally coupled \texorpdfstring{$f(R)$}{fR} gravity}\label{sec:solutions}

We are interested in solving Eq. (\ref{fried}) analytically. In order to do this we consider reasonable to assume a $\Lambda$CDM expansion, as in \cite{Dunsby:2010wg}, i.e.
\be\label{H}
H^2=\frac{\kappa^2}{3}\left(\frac{\rho_0}{a^3}+\Lambda\right).
\ee
Therefore, the scalar curvature reads
\be
R=\kappa^2\left(\frac{\rho_0}{a^3}+4\Lambda\right)\label{R}.
\ee
Consequently,
\be
a=\left(\frac{\rho_0}{R/\kappa^2-4\Lambda}\right)^{1/3},\label{a}
\ee
and
\be
3H^2=R-3\kappa^2\Lambda.\label{h2}
\ee
As we have two undetermined functions the way we are going to proceed is to specify one of them and solve for the other one. We are going to study the two simplest cases which seem interesting to us:
\begin{itemize}
 \item Set $f_2=R$ and find $f_1$.
 \item Set $f_1=R-2\Lambda \kappa^2$ and find $f_2$.
\end{itemize}
We also take $\mathcal{L}=-\rho$, so the energy-momentum tensor is conserved (see Eq. (\ref{conservation})), as already stated above. In addition, in this work, we will study two different cases of fluids: dust and a general perfect fluid.
\subsection{Finding $f_1$ with $f_2=R$}
Inserting Eqs. (\ref{R}) - (\ref{h2}) in Eq. (\ref{fried}) and letting $f_1$ as an undetermined function of $R$ and $f_2=R$ one obtains
\be\label{friedf11}
a_2(R)f_1'' +a_1(R)f_1'+a_0f_1=\frac{\kappa^2}{3}\left(\rho+\rho_{\lambda}\right),
\ee
where the prime stands for the derivative with respect to $R$ and
\bea
a_2(R)&\equiv&-\left(R-3 \kappa^2 \Lambda \right) \left(R-4 \kappa^2 \Lambda \right),\\
a_1(R)&\equiv&\left(\frac{R}{6}- \kappa^2 \Lambda \right),\\
a_0	&\equiv&\frac{1}{6},\\
\rho_{\lambda}&\equiv& 2\lambda(R-3\kappa^2\Lambda)\left(3\frac{d\rho}{dR}(R-4\kappa^2\Lambda)+\rho\right).
\eea
The first step is to find the homogeneous solution of the differential equation (\ref{friedf11}). We can see that the homogeneous part fo Eq.(\ref{friedf11}) is an hypergeometric differential equation and, in fact, it coincides with the differential equation obtained in minimal $f(R)$ gravity, which has been already computed in the literature. In \cite{Dunsby:2010wg} it was wrongly stated that there cannot be
any real valued function of Ricci scalar that can mimic a
$\Lambda$CDM expansion for a vacuum universe but one can see the alternative approach in \cite{homogeneous}, where an exhaustive analysis of the solutions is performed. Just notice that these solutions are given in terms of hypergeometric functions, so the convergence and integral representation of these functions in the physically possible regions must be adequately studied. In our case we are interested in the late-universe Cosmology, and the solution can be written as \cite{abramowitz}
\bea
f_1&=&c_0\left(\frac{\Lambda}{R-4\Lambda}\right)^{p_+-1} \\
&\times&{_2F_1}\left(q_+,p_+-1;r_+;-\frac{-\Lambda}{R-4\Lambda}\right), \nonumber
\eea
where $c_0$, $p_+$, $q_+$ and $r_+$ are real constants and $c_0\equiv-\bar{\omega}_1$, according to the notation in \cite{homogeneous}. The fact of having just one independent solution with the integration constant $c_0$ is due to the divergence that appears in the other linearly independent solution when $R\rightarrow\infty$. To avoid this troublesome issue, we set to zero the other integration constant. See \cite{homogeneous} for more details.
As we can notice, the terms including $\rho$ are exclusively on the right hand side of the differential equation (\ref{friedf11}), so the homogeneous solution is the same no matter the chosen content but the particular solution will change. We have calculated the particular solution for the two different cases stated above: dust and a general perfect fluid with a constant equation of state (EoS).
\subsubsection{Dust: $\rho=\rho_m$}
The first scenario we consider is a universe which contains just ordinary baryionic matter and cold dark matter.
In this case the matter density can be writen as
\be
\rho =\frac{R}{k^2}-4 \Lambda.
\ee
Inserting this into Eq. (\ref{friedf11}) one gets
\begin{eqnarray}\label{hyp}
a_2(R)f_1'' &+&a_1(R)f_1'+a_0f_1= \\
&&\frac{4}{3}\lambda a_2(R)+\frac{1}{3}\left(R-4 \kappa^2 \Lambda \right). \nonumber
\end{eqnarray}

One can compute a particular solution by proposing a second order polynomial on $R$.
Then one finds
\be\label{dust}
f_1=2\Lambda\kappa^2 \left(4 \kappa^2 \Lambda\lambda -1\right)+\left(1-4 \kappa^2 \Lambda \lambda\right)R+\frac{8}{9}\lambda R^2.
\ee
If $\lambda=0$, we recover GR: $f_1=R-2\Lambda \kappa^2$. We see that by considering a vanishing cosmological constant $\Lambda$ and a constant coupling the matter and the curvature term induces in this case a particular solution $f_1$ which is similar to Starobinsky inflationary model \cite{1980PhLB}, eventhough this latter scenario refers to the early universe. In this particular case $\lambda$ is proportional to the square of the scalaron mass.

\subsubsection{Perfect fluid with constant EoS: $\rho=\rho_w$}
Now we consider that the geometry still corresponds to a $\Lambda$CDM universe but filled with a perfect fluid with constant EoS $w$; i.e. its energy density reads:
\be
\rho=\frac{\rho_0}{a^{3(1+w)}}.
\ee
Therefore, by recalling Eq. (\ref{a}), we obtain
\be
\rho=\text{$\rho_0$} \left(\frac{\text{$\rho_0$}}{\frac{R}{\kappa^2}-4 \Lambda }\right)^{-(1+w)}.
\ee
Then, the Friedmann equation reads
\begin{eqnarray}\label{ww}
a_2(R)f_1'' &+&a_1(R)f_1' +a_0 f_1= \\
&-&\frac{1}{3}\rho_{0}^{-w}a_2(R)(R /\kappa^2-4 \Lambda)^{w} \nonumber \\
&&\left(\frac{1}{R-3\kappa^2\Lambda}+2\lambda-6(1+w)\lambda\right). \nonumber
\end{eqnarray}
If $w=0$, we recover Eq. (\ref{hyp}). We have not been able to find a particular solution for a general $w$ so we will
consider $w=-1/3$ as an example. This case is interesting as it lies at the boundary of the set
of matter fields that obey the
strong energy  condition.
In GR such fluids give rise to a
Milne  Universe
which
is a
coasting
universe, and the Ricci scalar is proportional to the square of the Hubble parameter. For this specific EoS one gets
\be\label{w}
f_1=a \left(R/\kappa^2-4  \Lambda \right)^{2/3}+b \left(R-3 \kappa^2 \Lambda \right) \left(R/\kappa^2-4 \Lambda \right)^{2/3},
\ee
where
\begin{align*}
 a&\equiv \frac{\kappa^2\sqrt[3]{\text{$\rho_0$}}}{3} \left(3 \kappa^2 \lambda  \Lambda +2 \right),\\
 b&\equiv \kappa^2 \lambda  \sqrt[3]{\text{$\rho_0$}}.
\end{align*}
Here we can check that by making $\lambda=0$ one recovers the prefect fluid solution in minimally coupled $f(R)$ theories given in \cite{Dunsby:2010wg}.

Finally, due to the linearity of the differential equation (\ref{ww}), as the coefficients $a_i$  do not depend on the fluid, it is easy to see that the particular solution for a universe containing both dust and a perfect fluid with EoS $w=-1/3$  is the sum of the previous particular solutions, i.e.  Eq. (\ref{dust}) and Eq. (\ref{w}).

\subsection{Finding $f_2$ with $f_1=R-2\kappa^2\Lambda$}

In this case we fix $f_1$ and let $f_2$ as an undetermined function. We have to solve the following differential equation
\begin{eqnarray}\label{friedf22}
6\lambda  &\Big[&\rho a_2(R) f_2'' -  \Big(\rho a_1(R) -  a_2(R)\frac{d\rho}{dR}\Big)f_2'-a_0\rho f_2\Big]= \nonumber \\
&&\rho+\Lambda-\frac{R-3\kappa^2\Lambda}{\kappa^2}. 
\end{eqnarray}
Let us notice that if $\lambda=0$, one obtains that the content of the universe can just be dust
\be
\rho=\frac{R}{\kappa^2}-4\Lambda.
\ee
This is consistent with Eq. (\ref{H}), as we are assuming a $\Lambda$CDM background.
Let us note that in this case $\rho$ cannot be collected as a simple inhomogeneous term in the differential equation (\ref{friedf22}). Then, in this case, both the homogeneous and particular solutions will be affected by the choice of $\rho$. We will solve this equation for dust and for a perfect fluid with a constant EoS.

\subsubsection{Dust: $\rho=\rho_m$}

The former differential equation (\ref{friedf22}) for a universe filled with baryonic matter and cold dark matter reads
\bea\label{dust1}
&6\lambda& \left(R-4 \kappa^2 \Lambda \right) \\ &\times&\Big( a_2(R) f_2'' +\left(\frac{5}{6} R-2 \kappa^2 \Lambda \right)f_2' - a_0f_2\Big)=0. \nonumber
\eea
As we can notice, the right hand side will be always zero no matter the coupling constant one chooses.
Assuming $\lambda\neq 0$ we note that we have an homogeneous differential equation with the following analytical solutions:
\begin{eqnarray}\label{pfw}
f_2&=&\frac{c_1}{\sqrt[3]{R-4 \kappa^2 \Lambda }}+\frac{6}{5} c_2 \sqrt{R-3 \kappa^2 \Lambda }\\
&+&\frac{9}{5} c_2 \sqrt{R-3 \kappa^2 \Lambda } \, _2F_1\left(\frac{5}{6},1;\frac{4}{3};4-\frac{R}{\kappa^2 \Lambda }\right), \nonumber
\end{eqnarray}
where $c_1$ and $c_2$ are integration constants and can be fixed by setting initial conditions on $f_2$ and $f_2'$. The integral representation of the hypergometric function is well defined in the range $4\kappa^2\Lambda\leq R< \infty$ \cite{abramowitz}, and as we are interested in the physical range, that is, $R\ge 4\kappa^2\Lambda$, this term is perfectly defined. One should have special care with the $c_1$ term, as in the limit $R\rightarrow 4\kappa^2\Lambda$ one would have zero at the denominator. However, given that this term appears in the action multiplied by the Lagrangian, it can be proven that the gravitational action is finite. Our solution is therefore perfectly defined.

\subsubsection{Perfect fluid with constant EoS: $\rho=\rho_w$}
\begin{figure}
\begin{center}
   \includegraphics[width=0.45\textwidth]{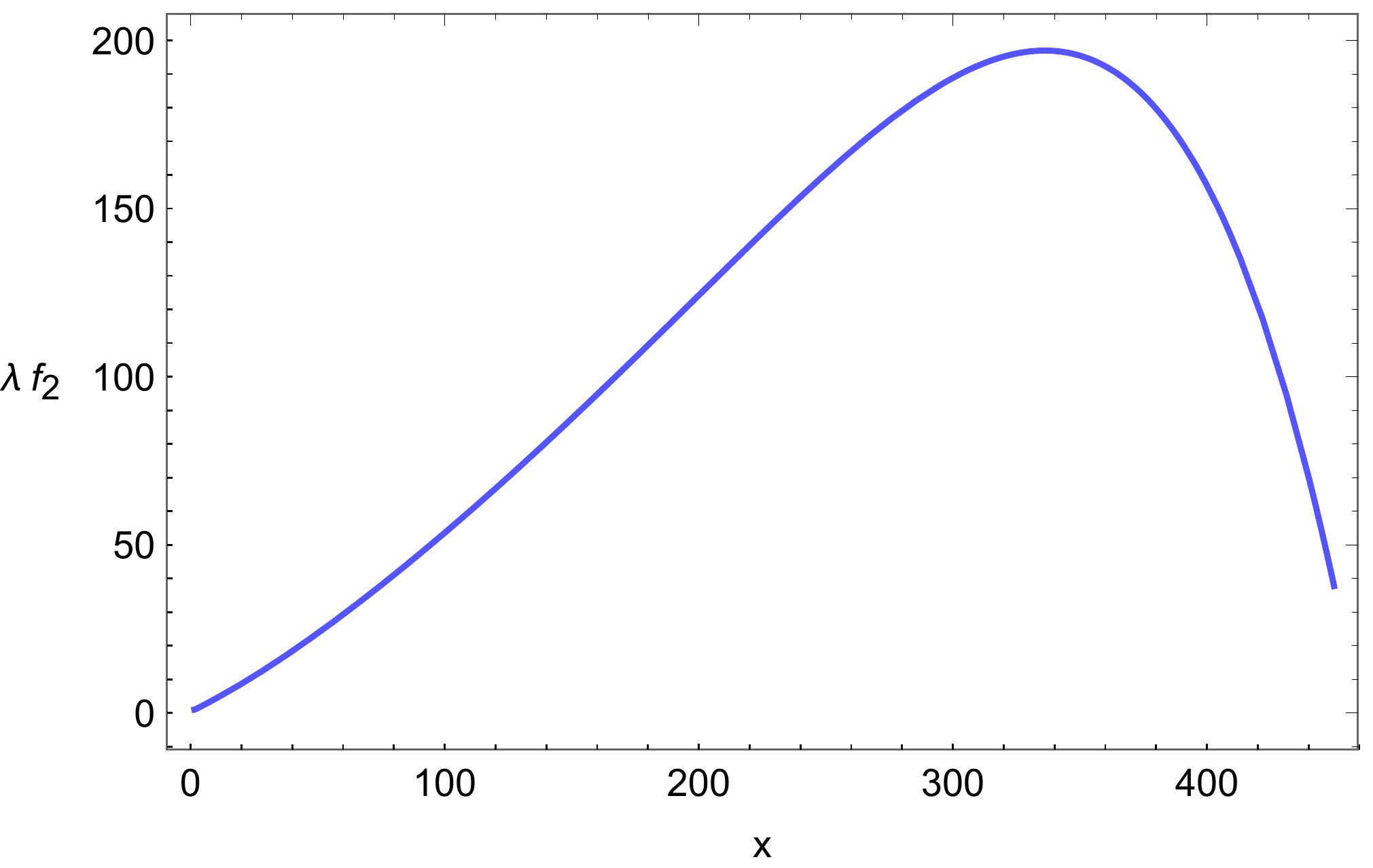}
   \caption{A numerical solution of Eq. (\ref{pf}) for $w=-1/3$ and $\lambda=10^{-5}$. We have set the initial conditions as $\lambda f_2(x=1.5)=1$ and $\lambda f_2'(x=1.5)=10^{-1}$. The solution covers the range from the present till $z=8.92$; i.e $x=450$. The vertical axis is the coupling $\lambda f_2$ while the horizontal axis is $x=\frac{R}{\kappa^2\Lambda}-3$.}
      \label{couplin}
\end{center}
\end{figure}

Now the equation which has to be solved is
\begin{eqnarray}\label{pf}
6&\lambda& \left[ a_2(R)f_2'' + \left[a_1(R) - (w+1) \left(R-3 \kappa^2 \Lambda \right) \right]f_2' \right. \nonumber \\
&&\left. + a_0 f_2\right]= -1+\left(\frac{\text{$\rho_0$}}{\frac{R}{\kappa^2}-4 \Lambda }\right)^w.
\end{eqnarray}
If $w=0$, one recovers, of course, the differential equation for a dust dominated universe, Eq. (\ref{dust1}). If $\lambda = 0$, one gets $\rho_0=R/\kappa^2-4\Lambda$, so this would be the only solution fixing $f_1=R-2\Lambda \kappa^2$.  Firstly, let us take a closer look at the homogeneous equation
\bea
\lambda && \left[ a_2(R)f_2''+ \Big(a_1(R) - (w+1) \left(R-3 \kappa^2 \Lambda \right)\Big)f_2'\right. \\
&&\left. + a_0f_2\right]=0.
\eea
As the right hand side does not depend on the coupling constant, the specific election of $\lambda$ will not affect the homogeneous solution. Assuming $\lambda\neq0$ and performing the change of variable
\be\label{xdef}
x=\frac{R}{k^2\Lambda}-3,
\ee
we can rewrite the previous equation as follows:
\be
(1-x) x f_2''+\left[\frac{1}{6}(x-3)-(w+1)x\right]f_2'+\frac{f_2}{6}=0.
\ee
We have found a solution which is valid for any value of $w$ which is given by:
\begin{eqnarray}\label{solpf}
f_{2,h}&=&C_1 \, _2F_1\left(a_1,b_1,c_1;\frac{R}{\kappa^2\Lambda}-3\right) \nonumber \\
&+& C_2 \left(\frac{R}{\kappa^2\Lambda}-3\right)^{7/6} _2F_1\left(a_2,b_2,c_2;\frac{R}{\kappa^2\Lambda}-3\right),
\end{eqnarray}
with
\begin{align}
 a_1&=w_1 -  \tilde{w},\\
 b_1&=w_1 +  \tilde{w},\\
 c_1&=-\frac{1}{6},\\
 a_2&=w_2 -  \tilde{w},\\
 b_2&=w_2+   \tilde{w},\\
 c_2&=\frac{13}{6}.
\end{align}
where we have defined $w_1\equiv-\frac{1}{12} + \frac{w}{2}$, $w_2\equiv\frac{13}{12} + \frac{w}{2}$ and $ \tilde{w}\equiv\frac{1}{12}\sqrt{25 - 12 w + 36 w^2}$.
Here we are giving the solution around the point $R=3k^2\Lambda$. As we stressed before, we are interested only in the physical region, that is, $R\ge 4\kappa^2\Lambda$, and in this region the variable  $x$ defined in Eq.  (\ref{xdef}), goes from 1 to $\infty$. The  condition for a well-defined integral representation in our case would apply to the region $(-\infty,1)$ \cite{abramowitz}. In order to have our solution defined within those limits, we change the variable to $1-x$, as it is set in \cite{abramowitz}. Rewriting the solution Eq. (\ref{solpf}), we have:
\bea
f_{2,h}&=&C_1 ~_2F_1\left(a_1,b_1,1+a_1+b_1-c_1;4-\frac{R}{\kappa^2\Lambda}\right)\\
&+&
C_2 \left(4-\frac{R}{\kappa^2\Lambda}\right)^{-w} \nonumber \\ &_2F_1&\left(c_2-a_2,c_2-b_2,1+c_2-a_2-b_2;4-\frac{R}{\kappa^2\Lambda}\right). \nonumber
\eea
The following step would be to find an analytical particular solution. In order to find a particular solution one would have to solve Eq. (\ref{pf}).
The problem is that it is not straight forward to find an analytical solution for this equation, not even for a specific $w$, however one can always find it by solving it numerically. In Fig. (\ref{couplin}) we show a numerical example of the total solution of Eq. (\ref{pf}). We have set $w=-1/3$ and $\lambda=10^{-5}$. Besides, we have chosen small
positive values for the initial conditions of $f_2$ and $f_2'$ so that the coupling does
not take too large values and does not change its sign.
The example we present here  is valid for the range which goes from
today until $z \simeq 9$.

\section{Cosmography within non-minimally coupled \texorpdfstring{$f(R)$}{fR} gravity}\label{sec:cosmogra}

In this section, we apply the cosmographic approach described in \cite{Visser:2004bf, Capozziello:2008qc, Cattoen:2007sk, Cattoen:2008th, BouhmadiLopez:2010pp, Capozziello:2011tj, Xia:2011iv, Morais:2015ooa} and adapt it to non-minimally coupled $f(R)$ gravity. In order to reduce the arbitrariness and complete our study we have considered the same two scenarios discussed before. In the first one, we set $f_2(R)=R$ so we can rename $f\equiv f_1$ for simplicity. In the second one, we will consider $f_1(R)=R-2\kappa^2\Lambda$ renaming $f\equiv f_2$ when doing the computations. We will consider that the universe is filled by dust. The procedure we will  follow is the one described, for example, in \cite{Morais:2015ooa}, where firstly one has to write the scalar curvature and derivatives in terms of the cosmographic parameters:
\bea\label{1}
R &=& 6 H^2 (1 - q),\\
\dot{R} &=& 6 H^3 (j - q - 2),\\
\ddot{R} &=&6 H^4 (s + q^2 + 8 q + 6),\\\label{4}
\dddot{R} &=& 6 H^5 (l - s - 2 (q + 4) j - 6 (3 q + 8) q - 24).
\eea
\subsection{Fixing $f_1$ for $f_2=R$}\label{subsec:cosmogra1}
The first step is to use the modified Friedman Eq. (\ref{fried}) and Raychaudhury equations to obtain expressions for $f$ and $f_{RRR}$. We obtain the Raychaudhury equation by taking the derivative of the Friedmann equation, Eq. (\ref{fried}) and the generalised conservation equation \ref{cons}. We rewrite Friedmann and Raychadhury equations as follows:
\bea\label{fried2}
f&=&\kappa^22\rho+(R-6H^2)f_R-6H\dot{R}f_{RR} \nonumber \\
&-&24 H^2\lambda\kappa^2\rho,\\\label{raych2}
f_{RRR}&=&-\frac{1}{\dot{R}^2}\left[\kappa^2\rho+2\dot{H}f_R+(\ddot{R}-H\dot{R})f_{RR} \right.\nonumber \\
&+&\left. \lambda\kappa^2\rho(8\dot{H}-12H^2)\right]. 
\eea
Differentiating the Raychaudhury equation with respect to the cosmic time, we obtain the third equation that we need:
\bea\label{third}\nonumber
\ddot{H}&=&\frac{\kappa^2}{2}\left[3H+\dot{R}\frac{f_{RR}}{f_R}\right]\frac{\rho}{f_R}-\frac{\dddot{R}-\dot{H}\dot{R}-H\ddot{R}}{2}\frac{f_{RR}}{f_R} \nonumber \\
&+&\frac{\ddot{R}\dot{R}-H\dot{R}^2}{2}\left(\frac{f_{RR}}{f_R}\right)^2\nonumber \\
&-&\frac{3\ddot{R}\dot{R}-H\dot{R}^2}{2}\frac{f_{RRR}}{f_{R}}+\frac{\dot{R}^3}{2}\frac{f_{RRR}f_{RR}}{f_R^2} \nonumber \\
&-&\frac{\dot{R}^3}{2}\frac{f_{RRRR}}{f_R} 
- \frac{\lambda\kappa^2\rho}{2f_R}\left(g_1(t)+\frac{f_{RR}}{f_R}g_2(t)\right),
\eea
where
\bea
g_1(t)&=&36H^3-48H\dot{H}+8\ddot{H},\\
g_2(t)&=&12H^2\dot{R}-8\dot{H}\dot{R}.
\eea
Since at  present we expect $f(R)$ to  not deviate too much from GR, we can approximate our theory by its Taylor expansion around $R_0$ up to second order\footnote{Although in reference \cite{Morais:2015ooa}, the authors  carried the analysis up to third order, we carry here the expansion only up to $O(R-R_0)^3$. This way we do not need to fix $f_R=1$.}, being $R_0$ the present value of the scalar curvature:
\bea
f(R)&\simeq& f(R_0)+f_R(R_0)(R-R_0) \\
&+&\frac{f_{RR}(R_0)}{2}(R-R_0)^2+O(R-R_0)^3. \nonumber
\eea
Then, we can ignore the terms containing $f_{RRR}$ and $f_{RRRR}$ in Eq. (\ref{third}).
Now we have three equations and we can express $f$, $f_R$ and $f_{RR}$ in terms of the cosmographic parameters using Eqs. (\ref{1})-(\ref{4}) and
\bea\label{hd}
\dot{H} &=& -H^2 (1 + q),\\\label{hdd}
\ddot{H} &=& H^3 (j + 3 q + 2),
\eea
which come from the basic formulas of cosmography. Please note that from now on, $\Omega_m$ and all the cosmographic parameters refers to quantities evaluated today. The cosmographic adimensional quantities we compute are given by
\bea\label{cosmoparam1}
\frac{f(R_0)}{6H_0^2}=\frac{A_0\Omega_{m}+\lambda H^2_0\Omega_{m} C_0}{D},\\
f_{R}(R_0)=\frac{A_1\Omega_{m}+\lambda H^2_0\Omega_{m}  C_1}{D},\\\label{cosmoparam11}
\frac{f_{RR}(R_0)}{(6H_0^2)^{-1}}=\frac{A_2\Omega_{m}+\lambda H^2_0\Omega_{m}  C_2}{D},
\eea
where the explicit expression of the coefficients $A_i, C_i$ with $i=1,2,3$ and $D$ are shown in the Appendix.

We can get an order of magnitude of the cosmographic parameters by computing them for
a given dark energy phenomenological parameterisation.
The best and simplest one is the $\Lambda$CDM model \cite{Capozziello:2008qc},
\bea
j&=&1,\\
q&=&\frac{3 \text{$\Omega_{m}$}}{2}-1,\\
s&=&1 -\frac{9 \text{$\Omega_{m}$}}{2}.
\eea
Afterwards, we took the latest \textit{Planck} results \cite{Ade:2015xua} to compute the cosmographic parameters choosing a specific value for\footnote{Despite not having quantitative information about the order of magnitude of the coupling parameter, we estimate it to be sufficiently small so that it does not produce a drastic deviation from  Einstein-Hilbert action.} $\lambda$ :
\bea
\Omega_{m}&=&0.315\, , \nonumber\\ 
q&=&-0.540\, , \nonumber \\
j&=& 1\, , \nonumber \\ 
s&=&-0.379\, ,\nonumber \\
\lambda H_0^2&=&\frac{1}{10^{15}}. 
\eea
One obtains:
\begin{eqnarray}\label{adim1}
\frac{f(R_0)}{6H_0^2}&=&0.84675  \\
f_{R}(R_0)&=&1, \\ \frac{f_{RR}(R_0)}{(6H_0^2)^{-1}}&=&3.98651\cdot 10^{-15}.
\end{eqnarray}
It can be shown that if the model is close enough to $\Lambda$CDM, then $f$ and $f_R$ are decreasing functions of $\lambda$. We have assumed the latest \textit{Planck} data \cite{Ade:2015xua}. Likewise, it can be shown that $f_{RR}$ is an increasing function of $\lambda$.  Consequently, and under the assumption $f_2(R)=R$, this result can be interpreted as follows: the larger is the non-minimal coupling in the action Eq. (\ref{action}), the smaller is the pure gravitational part; i.e. $f_1(R)$.
\subsection{Fixing $f_2$ for $f_1=R-2\kappa^2\Lambda$}

Following a procedure similar to the one used previously, i.e. starting from the Friedmann equation and getting the Raychaudhury equation, we obtain
\bea\label{friedf1}
 \kappa^2  \lambda  \rho f &=&  \kappa^2
 \lambda  \rho(12  H^2  + R)f_R \\ 
&-&6 \kappa  \lambda  \rho  H  \dot{R} f_{RR} +3 H^2-\kappa^2  \Lambda -\kappa^2 \rho, \nonumber  \\\label{raychf1}
2 \kappa^2  \lambda  \rho  \dot{R}^2f_{RRR}  &=& \kappa^2  \rho  (1+f \lambda)+ 
\kappa^2 \lambda \rho \left(2 \dot{H} -24 H^2\right) f_{R} \nonumber \\   
&+& \kappa^2 \lambda \rho (14 H   \ddot{R}-2 \ddot{R}) f_{RR}   +2  \dot{H}.
\eea
One can note that if one sets $\lambda=0$, one gets $3H^2=\kappa^2 (\rho+\Lambda)$ from the Friedmann equation and $2\dot{H}=-\kappa^2\rho$ from the Raychaudhury equation. This is completely consistent with the background we are considering (please c.f. Eq. (\ref{H})).

We obtain a third equation deriving the Raychaudhury equation with respect to the cosmic time,

\bea\label{thirdf1}\nonumber
&-6& \left(H \ddot{H}+\dot{H}^2\right)=3 \kappa^2  \rho(\dot{H}-3 H^2) \\
&&+\kappa^2  \lambda  \rho\Big[\left(-9 H^2  +3  \dot{H}\right)f \nonumber \\
&&+ \left(216 H^4   -162  H^2 \dot{H}  +24  H \ddot{H}  +6  \dot{H}^2 \right)f_{R}\Big. \nonumber \\
&&+ \left(-198 H^3   \dot{R}
 +60 H^2  \ddot{R}+66 H \dot{H}   \dot{R}-6 H   \dddot{R} \right.\nonumber \\
&& \left. -6 \dot{H}   \ddot{R}+ \dot{R} (\dot{R}-6 \ddot{H})\right) f_{RR} \nonumber \\
&&\Big.+ \left(60  H^2   \dot{R}^2-18  H  \dot{R} \ddot{R}-6 \dot{H}  \dot{R}^2\right)f_{RRR} \nonumber \\
&&-6   \dot{R}^3 H f_{RRRR}\Big].
\eea

As explained before, we can ignore the terms containing $f_{RRRR}$ and $f_{RRR}$. Using an approach analogous to the one in Subsec. \ref{subsec:cosmogra1}, we can express $f$, $f_R$ and $f_{RR}$ in terms of the cosmographic parameters considering baryonic and dark matter as the only density content. Again, note that from now on, $\Omega_m$ and all the cosmographic parameters refers to quantities evaluated at present. Similarly to what we did in the previous subsection, we compute the following adimensional quantities:
\bea\label{adim}
\lambda \Omega_{m}f(R_0)=\frac{A_0 \Omega_{m}+B_0+ C_0\frac{\kappa^2\Lambda}{H_0^2}}{3 D},\\
\lambda \Omega_{m}\frac{f_{R}(R_0)}{(6 H_0^2)^{-1}}=\frac{A_1\Omega_{m}+B_1+ C_1\frac{\kappa^2\Lambda}{H_0^2}}{ D},\\\label{adim3}
\lambda \Omega_{m}\frac{f_{RR}(R_0)}{(6 H_0^2)^{-2}}=\frac{A_2 \Omega_{m}+B_2+ C_2\frac{\kappa^2\Lambda}{H_0^2}}{D},
\eea
where the values of the coefficients appear explicitely in the Appendix.

\begin{figure}
\begin{center}
    \includegraphics[width=0.45\textwidth]{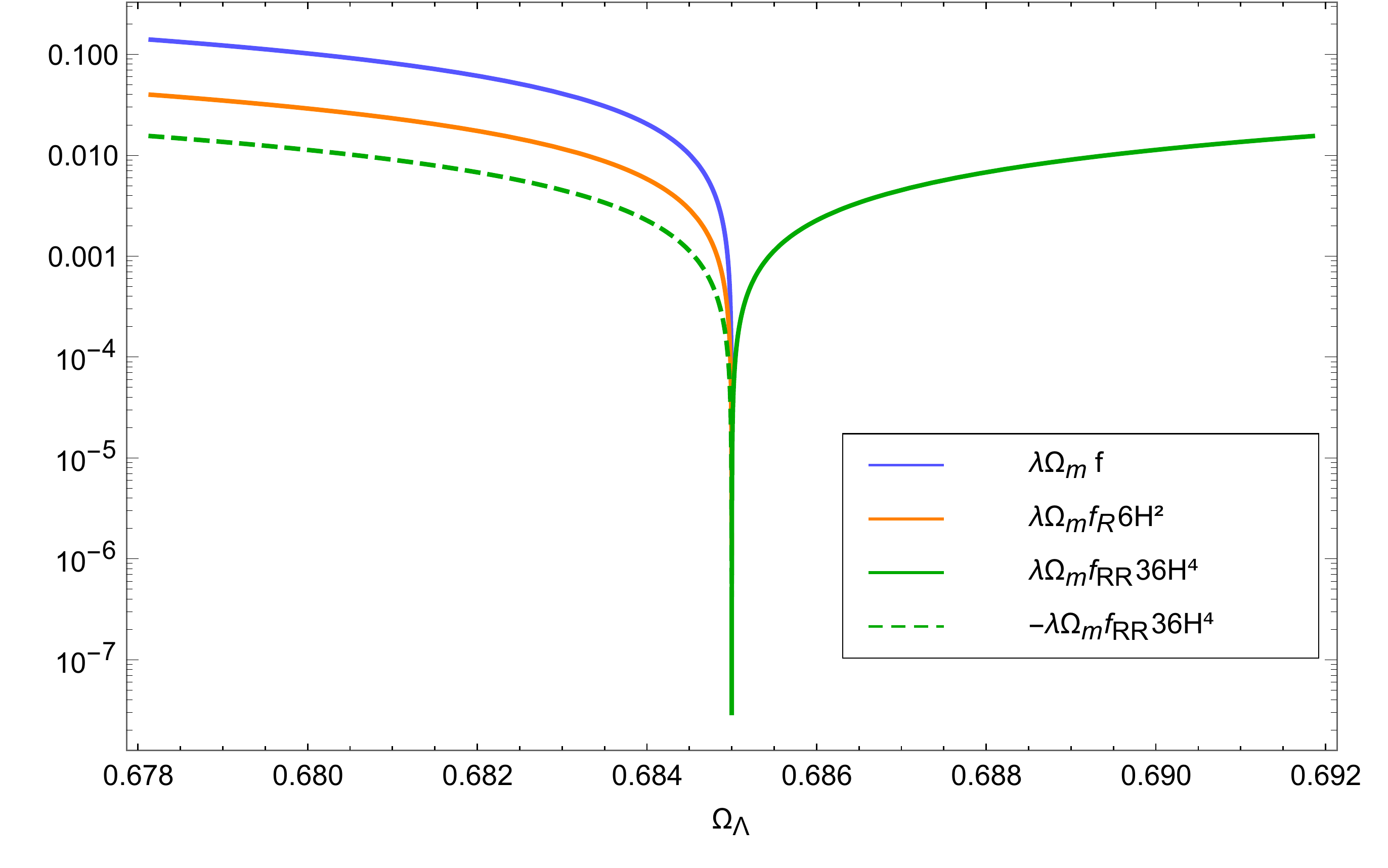}
  \caption{Here the adimensional quantities given in Eqs.  (\ref{adim})-(\ref{adim3}) are represented as a function of $\Omega_{\Lambda}$. 
}
\label{omegalambda}
\end{center}
\end{figure}

Computing the parameters for a $\Lambda$CDM model,
one gets
\begin{eqnarray}
\lambda \Omega_{m}f(R_0)&=&0, \\
\lambda \Omega_{m}\frac{f_{R}(R_0)}{(6H_0^2)^{-1}}&=&0, \\
\lambda \Omega_{m}\frac{f_{RR}(R_0)}{(6H_0^2)^{-2}}&=&0.
\end{eqnarray}
This result is completely consistent with our model. Having a very carefull look at Eq. (\ref{friedf1}) one can see that the only way to have a $\Lambda$CDM background is when $\lambda$ is exactly zero. Nevertheless, we are interested in having a small non-null coupling constant. This means that we have to relax the condition $\Omega_{\Lambda}=1-\Omega_m$ and leave $\Omega_{\Lambda}$ unfixed. The value of the coupling constant would affect the value of the density of the cosmological constant. In order to see it qualitatively we draw Fig. (\ref{omegalambda}). It is straight forward to notice that in the limit $\lambda\rightarrow 0$, the value of $\Omega_{\Lambda}$ tends to exactly $1-\Omega_m$. Something interesting that one can also observe is that the larger is the coupling constant $\lambda$, the smaller is the dark energy density $\Omega_{\Lambda}$. This is a very desirable feature, as one of the motivations of introducing a coupling to matter is that such a coupling could alleviate the dark energy puzzle.

\section{Results}\label{sec:results}
In this work, we have considered a modified gravity model characterised by having a non-minimal coupling to matter. A good representative of this kind of models was first proposed in \cite{extraforce} and it corresponds to a metric gravity extension of the well known $f(R)$ gravity. Within this framework, we have two functions: $f_1(R)$, uncoupled to matter and generalising  Hilbert-Einstein action, and $f_2(R)$, which takes into account  the coupling of gravity to matter (see Eq. (\ref{action})). Our main interest resides on getting exact analytical solutions for this theory in the context of the late-time Cosmology. For this purpose, we have selected  two simple and physically meaningful scenarios; i.e. we have obtained two $f(R)$ non-minimally coupled models that mimic a $\Lambda$CDM setup.

For the first case, we have chosen $f_2(R)=R$, i.e. a small coupling of gravity to matter. Then, $f_1(R)$ has to satisfy Eq. (\ref{friedf11}). Solving this equation, we have obtained a physically meaninful solution which exactly mimics a $\Lambda$CDM expansion for a vacuum universe, which is in contrast with what it is claimed in \cite{Dunsby:2010wg}. Then, we have considered that the universe is filled up with an energy-density content and we have analysed two different scenarios. For a matter-dominated universe, the solution is described by the usual $R-2\kappa^2\Lambda$ plus a polynomical correction on $R$ driven by the coupling constant, $\lambda$. It is interesting to notice how in this solution it naturally appears a term which acts as an effective cosmological constant.
For a general perfect fluid-dominated universe with a constant EoS, we have been able to find an analytical solution once we have chosen the specific EoS $w=-1/3$, which can be interpreted as an energy-density mimicking an open universe. The obtained solution is an algebraic expresion depending on $R$ and it is given in Eq. (\ref{w}).

For the second case, we have chosen $f_1(R)=R-2\kappa^2\Lambda$. Then $f_2(R)$ has to satisfy Eq. (\ref{friedf22}). We have analysed the two setups previously considered. For the the matter-dominated universe, the solution we have found is given by a hypergeometric function and includes some algebraic corrections on $R$ (see Eq. (\ref{pfw})). We have carried a detailed analysis ensuring the convergence of the Lagrangian in the range which is physically meaningful. For the general perfect fluid-dominated universe, we have been able to find an homogeneous solution which, in fact, is valid for every constant EoS, $w$. However, we have not been able to find an analytical particular solution, though, it is possible to get it numerically. In Fig. (\ref{couplin}) we present an example.

Finally, to complete our study and extract further information about the two analysed models, we have performed a cosmographic analysis. For the first case, we have obtained that the quantities $f$ and $f_R$ are decreasing functions of $\lambda$ and this can be interpreted as follows ``the stronger is the coupling to matter, the weaker is the weight of the pure gravitational part of the action''. For the second case, the obtained result leads us to the conclusion  that the only choice which allows for an exact $\Lambda$CDM background corresponds to $\lambda=0$. Then, in order to allow a non-zero value of $\lambda$, we have to let the parameter $\Omega_{\Lambda}$ free. The way $\lambda$ affects the value of $\Omega_{\Lambda}$ shows a desirable result because the larger is $\lambda$, the smaller is $\Omega_{\Lambda}$. This satisfies the main motivation of our work, as the non-minimally
coupled term can  mimic the role of a dynamical  dark energy, being responsible of
the late-time acceleration of the universe. 
Please, notice that the cosmographic reconstruction carried in our
work is one with respect to the scalar curvature; i.e. we obtain $f_1(R)$
and $f_2(R)$. An alternative approach is to carry out a cosmographic
reconstruction with respect to the redshift, i.e. obtaining $f_1(z)$
and $f_2(z)$. This has been carried out in \cite{Capozziello:2014zda} showing how $f$ modified
gravity can match the acceleration-deceleration transition in
agreement with the  large-scale structure formation. Of course, in our
case given that we do not have an algebraic expression for $H(z)$,
this would have to be tackled numerically. We leave this interesting
approach for a future work where we will not only tackle theoretical
issues as was done here but where we will further constrain our model
observationally.

\section*{Acknowledgments}

M.~B.-L., M.~O.~B. and R.~L. were supported by the
Spanish Ministry of Economy and Competitiveness through
research projects No. FIS2017-85076-P (MINECO/AEI/\\FEDER, UE)
and also by the Basque Government through research
project No. GIC17/116-IT956-16. M.O.B. acknowledges financial
support from the FPI grant BES-2015-071489. The work of M.B.L. is also supported by the Basque Foundation of Science
IKERBASQUE. This article is based upon work from COST Action CA15117
(CANTATA), supported by COST (European Cooperation
in Science and Technology).

\bibliographystyle{apsrev4-1}
\bibliography{biblio.bib}

\section*{Appendix}

We summarise the cosmographic approach we used in Sec.~\ref{sec:cosmogra}.
\subsection{Cosmography: Fixing $f_1(R)$ for $f_2=R$}

We started using Eqs. (\ref{fried2})-(\ref{third}), where we substitute $R$ and its derivatives using Eq. (\ref{1})-(\ref{4}). Then we carried a similar procedure by substituting the time derivative of the Hubble rate, $H$, using Eqs. (\ref{hd}) and (\ref{hdd}), on the resulting equations. Finally, we evaluated the three resulting equations at present time and deduced Eqs.  (\ref{cosmoparam1})-(\ref{cosmoparam11}), where
\begin{eqnarray}
A_0&=& -4 j^2-5 j q^2-18 j q-2 j s+24 j-15 q^3+6 q^2 \nonumber \\
&+&q\,s+60 q+12\, , \\
C_0&=&12 \bigg\{2 j^2 (q+2)+j \bigg[q \bigg(q (4 q+5)-2 (s+6)\bigg) \nonumber \\
&+& 2 (s-24)\bigg]+3 q
\bigg[q \left(6 q^2+q-12\right)-4\bigg] \nonumber \\
&-& q (14 q+13) s+36\bigg\}\, , \\
A_1 &=& 3 \left(3 j q+12 j+15 q^2+26 q-s+4\right),\\\nonumber
C_1&=& -12 \bigg[2 j^2+j \big(q (4 q+33)-2 s+54\big)  \nonumber \\
&+& q \left(18 q^2+9 q-14 s-48\right)-15 (s+4)\bigg]\, ,\\
A_2&=& 6(-1 + j)\, ,  \\
C_2&=& 72 \bigg[-j+4 q (q+2)+5\bigg]\, ,\\
D&=&2\bigg[j^2 + 3 (1 + q)^2 (4 + 5 q) \nonumber \\
&+& j \big(3 + 2 q (3 + q) - s\big) - q s\bigg]\, .
\end{eqnarray}

\subsection{Cosmography: Fixing $f_2(R)$ for $f_1=R-2\kappa^2\Lambda$}

Following a similar procedure to the one explained above we got Eqs. (\ref{adim})-(\ref{adim3}), where
\begin{eqnarray}
A_0&=&-12 \bigg\{4 j^2 + 3 (12 + s) - j \bigg[6 + q (6 + 7 q) + s\bigg]  \nonumber \\
&+& q \bigg[3 q (32 + 9 q) + 2 (51 + s)\bigg]\bigg\}\, , \\
B_0&=&-\bigg(-1062 - 12 j^2 (1 + 2 q) \nonumber \\
&+& 6 j \bigg\{-18 + q \bigg[-21 + q (38 + 7 q)\bigg]\bigg\} \nonumber \\ 
&-&9 q \bigg\{26 + q \bigg[-49 + q (69 + 22 q)\bigg]\bigg\} \nonumber \\
&+&6 (15 + j - 8 q) (-1 + q) s\bigg),\\
C_0&=&-\bigg\{978 + 28 j^2 + q \bigg[1146 + 9 q (21 + q) - 16 s\bigg] \nonumber \\
&+& 84 s -
  2 j (48 - 9 q + 5 q^2 + 2 s)\bigg\}\, ,\\
A_1 &=& 0\, ,\\
B_1 &=&-8 j^2+2 j \bigg[7 q (q+3)+s+12\bigg] \nonumber \\
&-&q \bigg[3 q (22 q+83)+16 (s+6)\bigg]+6 (s+13),\\
C_1&=& -33 q^2 + 6 j (2 + q) - 2 (53 + 70 q + 4 s),\\
A_2&=&0\, ,\\
B_2&=& -12\bigg[j-2 q (2 q+1)+1\bigg],\\
C_2&=& 4(5 + j + 6 q)\, , \\\nonumber
D&=&4 \bigg\{4 j^2 + 3 (12 + s) - j \bigg(6 + q (6 + 7 q) + s\bigg) \nonumber \\
&+&  q \bigg[3 q (32 + 9 q) + 2 (51 + s)\bigg]\bigg\}\, .
\end{eqnarray}

\end{document}